\documentclass[
reprint,twocolumns,
superscriptaddress,
amsmath,amssymb,
aps,
nofootinbib,
prb,
]{revtex4-2}

\usepackage{booktabs}

\usepackage{amsfonts}

\usepackage{algorithm}
\usepackage{algpseudocode}

\usepackage{graphicx}
\graphicspath{{./figures/}}
\usepackage{dcolumn}
\usepackage{bm}
\usepackage{hyperref}
\usepackage{cleveref}
\usepackage{physics}
\usepackage{xcolor}

\usepackage{mathrsfs}
\usepackage{mathtools}
\usepackage{comment}
\usepackage{dsfont}
\usepackage{euscript}

\hypersetup{colorlinks=true, citecolor=purple, linkcolor=purple, urlcolor=purple}

\begin{document}

\title{Classically Prepared, Quantumly Evolved: Hybrid Algorithm for Molecular Spectra}

\author{Alessandro Santini}
\affiliation{CPHT, CNRS, Ecole Polytechnique, Institut Polytechnique de Paris, 91120 Palaiseau, France}
\affiliation{Collège de France, Université PSL, 11 place Marcelin Berthelot, 75005 Paris, France}
\affiliation{Inria Paris-Saclay, Bâtiment Alan Turing, 1, rue Honoré d’Estienne d’Orves – 91120 Palaiseau}
\affiliation{LIX, CNRS, École polytechnique, Institut Polytechnique de Paris, 91120 Palaiseau, France}

\author{Stefano Barison}
\thanks{Current address: \\ \textit{IBM Quantum, IBM Research Europe - Zürich, CH-8803 Rüschlikon, Switzerland} \\
\textit{Institute for Theoretical Physics, ETH Zürich, CH-8093 Zürich, Switzerland}}
\affiliation{Institute of Physics, Ecole Polytechnique Fédérale de Lausanne (EPFL), CH-1015 Lausanne, Switzerland}

\author{Filippo Vicentini}
\affiliation{CPHT, CNRS, Ecole Polytechnique, Institut Polytechnique de Paris, 91120 Palaiseau, France}
\affiliation{Collège de France, Université PSL, 11 place Marcelin Berthelot, 75005 Paris, France}
\affiliation{Inria Paris-Saclay, Bâtiment Alan Turing, 1, rue Honoré d’Estienne d’Orves – 91120 Palaiseau}
\affiliation{LIX, CNRS, École polytechnique, Institut Polytechnique de Paris, 91120 Palaiseau, France}

\begin{abstract}
We introduce a hybrid classical–quantum algorithm to compute dynamical correlation functions and excitation spectra in many-body quantum systems, with a focus on molecular systems. The method combines classical preparation of a perturbed ground state with short-time quantum evolution of product states sampled from it. The resulting quantum samples define an effective subspace of the Hilbert space, onto which the Hamiltonian is projected to enable efficient classical simulation of long-time dynamics. This subspace-based approach achieves high-resolution spectral reconstruction using shallow circuits and few samples. Benchmarks on molecular systems show excellent agreement with exact diagonalization and demonstrate access to dynamical timescales beyond the reach of purely classical methods, highlighting its suitability for near-term and early fault-tolerant quantum hardware.
\end{abstract}

\maketitle

\section{Introduction}

Dynamical observables are central to quantum chemistry and condensed matter physics, yet are markedly harder to compute than static ground-state quantities.
For example, spectral response functions encoding excitation energies, intensities, and lifetimes, can be directly compared to experiments like ARPES~\cite{Damascelli2003}. 
Moreover, they are the central object of Dynamical Mean Field Theory, often used to treat realistic multi-band materials~\cite{Kotliar2006}.

For Quantum Monte Carlo and diagrammatic methods, accessing the low-temperature regime is challenging because of instabilities in the analytic continuation from the imaginary plane or because of the sign problem.
Classical methods that are highly accurate for ground or low-lying excited states such as Variational Monte Carlo (VMC~\cite{sorella}) and tensor-network algorithms like Density Matrix Renormalization Group (DMRG~\cite{dmrg_1992,DMRGageof,DMRGX,Collura_2024}) have been extended to the realm of real-time dynamics both for VMC~\cite{MendesSantos2023Spectral,Gravina2025Systematic,Sinibaldi2024Galerkin,Walle2024tNQS} and TNs~\cite{Krinitsin2025ttn,Schollwock_2011,Schollwock_2019,TDVP_2011,TDVP_2016}.
Nevertheless, several issues from increasing sampling complexity~\cite{Sinibaldi2023unbiasing} and growing entanglement~\cite{Schollwock_2011,Schollwock_2019,Calabrese_2005} in the presence of strong correlations or orbital degeneracies~\cite{Foerster2024} make those simulations challenging and hard to scale.

Quantum computers are emerging as a promising route to simulate quantum many-body systems ~\cite{Feynman1982,Feynman85,lloyd1996,Tacchino_2019}, with early algorithms targeting chemistry and condensed-matter problems~\cite{Aspuru-Guzik2005,farhi2000adiabatic,Babbush2014,Libor2014}.
However, near-term devices remain constrained by limited qubit count, restricted connectivity, and short coherence times~\cite{Reiher2017,childs2018_pnas,babbush2018_prx,yunseong2019_npjqi}, preventing their application to large-scale problems.
Approaches based on variational optimization of parametrized circuits ~\cite{Peruzzo2014,Kandala2017,cerezo2020variational,Motta_2019,2019_grimsley_adapt-vqe,2022_bharti_nisq_algo_rev,Benchen2022,Eddins_2022,barison2023embedding} by minimizing the ground-state energy in a similar spirit to VMC or DMRG attempted to circumvent those limitations.
However they were eventually revealed to be sensitive to statistical and hardware noise~\cite{Astrakhantsev2023} and require substantial sampling, making them uncompetitive for ground-state search against classical variational methods in many regimes~\cite{Mazzola2024}.

In contrast, while pushing real-time evolution to long times is notoriously difficult for classical algorithms, unitary time evolution (exactly or approximately) is a native quantum task~\cite{Trotter1959,1976_suzuki,lloyd1996,Berry_2015,Tacchino_2019}, and several techniques have been developed to render short-time dynamics feasible on contemporary quantum hardware~\cite{Ying2017rte,Campbell_2019,lin2020real,barison2021efficient,berthusen_2022_PhysRevX.8.011021dynamics_on_hardware,Miessen2023,Linteau2024}.

Complementary to this, quantum subspace methods~\cite{Colless2018_qse,Motta_2019,Takeshita2020,Kirby2023exactefficient,Tkachenko_2024,Kirby_2024,Motta_2024} project the Hamiltonian onto a low-dimensional, physically relevant subspace selected with the aid of a quantum device.
They are particularly effective when the physically relevant portion of the Hilbert space is small but difficult to identify classically.
Among these strategies, sample-based variants have recently pushed static-property calculations beyond classical heuristics~\cite{kanno2023qsci,nakagawa2023adaptqsci,RobledoMoreno2025,Barison_2025,shajan2024sqd,liepuoniute2024sqd,danilov2025sqd,yu2025sqkd,N_tzel_2025,mikkelsen2025qscite,piccinelli2025}.
Although generalizations to excited states exist~\cite{kanno2023qsci,Barison_2025}, they typically presume access to a low-depth approximation of the ground state, limiting applicability when entanglement is nontrivial.

In this manuscript we build upon those premises, and  investigate what efficient hybrid schemes could be built by combining classical variational methods with purely deterministic, nonparametrized quantum circuits and subspace diagonalization.
Our goal is to leverage the accuracy of classical ground-state optimization together with the low cost of short-time quantum dynamics on hardware.
We introduce a hybrid classical–quantum scheme for zero-temperature dynamical correlation functions that sidesteps ground-state preparation on hardware.

The manuscript is organized as follows.
\Cref{sec:methods} formalizes the estimator and the subspace-projection procedure for reconstructing the dynamical correlator $G_A(t)$; \Cref{sec:result} benchmarks molecular spectra against exact diagonalization and tensor-network references; \Cref{sec:conclusions} discusses implications for quantum hardware implementations and extensions.

\section{Methods}
\label{sec:methods}
We study molecular systems within the Born–Oppenheimer approximation \cite{Born1927}. 
By fixing the nuclear positions and choosing a suitable basis, the electronic Hamiltonian can be expressed as

\begin{equation}
\hat{H} = E_{\rm nuc} + \sum_{\substack{sr}} h_{pq}^{(1)}\,\hat{c}^\dagger_{p}\,\hat{c}_{q} + \frac{1}{2}\sum_{\substack{prqs}} h_{pq,sr}^{(2)}\,\hat{c}^\dagger_{p}\,\hat{c}^\dagger_{q}\,\hat{c}_{s}\,\hat{c}_{r}\,.
\label{eq:hamiltonian}
\end{equation}
Here the indices $p$, $r$, $q$, and $s$  run over both spatial and spin degrees of freedom. 
The spatial part is represented by a set of $M$ orthonormal functions, $\{ \phi_p \}_M$,  such that the total number of spin–orbitals is $2M$.
The constant $E_{\rm nuc}$ takes into account the energy offset of fixing the nuclei, $h^{(1)}_{pq}$ are the one-electron integrals, and $h_{pr,qs}^{(2)}$ are the two-electron integrals; all quantities are given in atomic units. 
The stationary states of $\hat H$ satisfy the time-independent Schr\"{o}dinger equation $\hat H\ket{\psi_n} = E_n\ket{\psi_n}$. We denote by $\ket{\psi_0}$ the normalized ground state, i.e. $\braket{\psi_0}$=1, with energy $E_0$.

Given an Hamiltonian $\hat{H}$, its ground state $\ket{\psi_0}$ and an operator $\hat{A}$, we aim to compute the zero-temperature dynamical correlation
\begin{equation}
\label{eq:corr_initial}
G_{A}(t) = \langle \psi_0 | \hat{A}^\dagger(t) \hat{A} | \psi_0 \rangle , 
\end{equation}
where the time-evolved operator in the Heisenberg picture is $\hat{A}^\dagger(t) = \hat{U}^\dagger(t) \hat{A}^\dagger \hat{U}(t)$, with $\hat{U}(t) = e^{-i\hat{H}t}$. This quantity describes how a perturbation $\hat A$ propagates through the system in real time.
Rewriting the expression, we obtain
\begin{align}
G_A(t) &= e^{iE_0t}\langle \psi_0 | \hat{A}^\dagger \hat{U}(t) \hat{A} | \psi_0 \rangle \\\notag
&= e^{iE_0t}\, \langle \hat{A}^\dagger \hat{A} \rangle_0\, \langle \psi_A | \hat{U}(t) | \psi_A \rangle\,,
\end{align}
with $\langle \hat{A}^\dagger \hat{A} \rangle_0 = \langle \psi_0 | \hat{A}^\dagger \hat{A} | \psi_0 \rangle$ and the normalized perturbed state $\ket{\psi_A} = \frac{\hat{A}\ket{\psi_0}}{\sqrt{\langle \hat{A}^\dagger \hat{A} \rangle_0}}$. 
The Fourier transform
\begin{align}
\tilde{G}_A(\omega) &= \int_{-\infty}^{\infty} dt\, e^{i\omega t} G_A(t)\,\\\notag
&= \langle \hat{A}^\dagger \hat{A} \rangle_0 \sum_n \delta[\omega - (E_n - E_0)]\, |\braket{\psi_A}{E_n}|^2,
\end{align}
probes the spectral decomposition of $\ket{\psi_A}$, highlighting the transition probabilities to the excited states that contribute most significantly to the dynamics.

\begin{figure*}
    \centering
    \includegraphics[width=\linewidth]{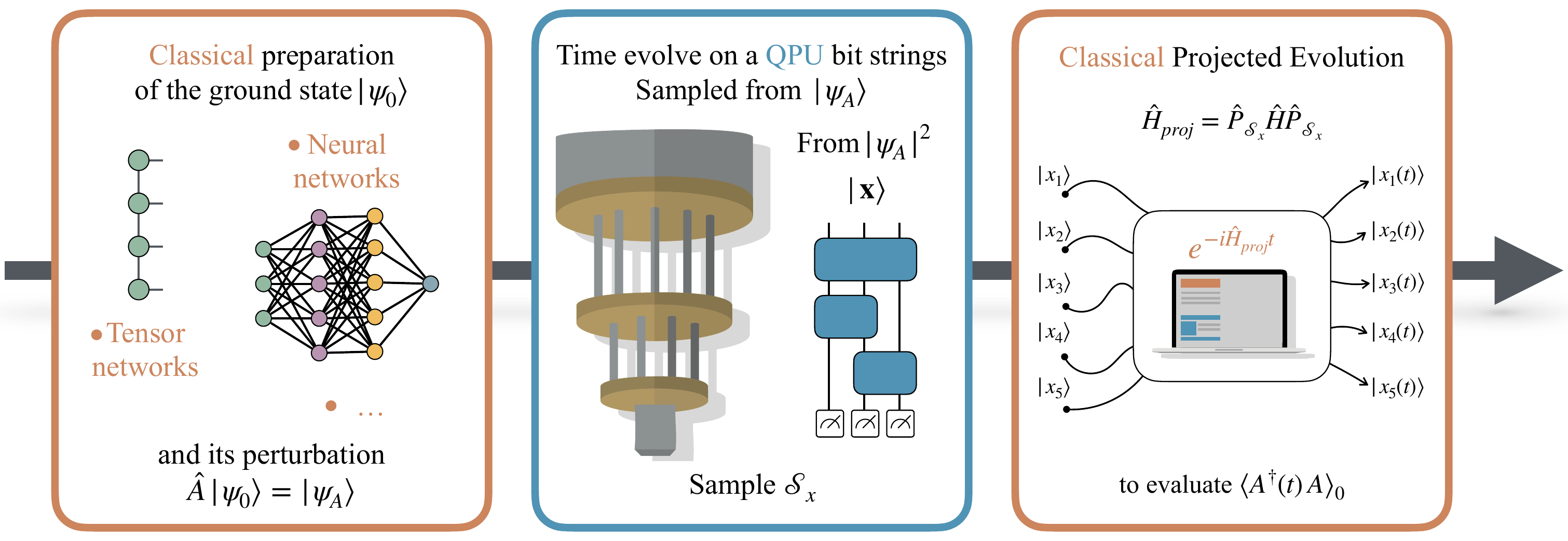}
    \caption{\textbf{Hybrid classical–quantum workflow.}
A classical approximation of the perturbed state $\ket{\psi_A\rangle}$ is sampled to generate product states, which are briefly evolved on quantum hardware.
The measured outcomes define small subspaces $\mathcal{S}_x$ onto which the Hamiltonian is projected for efficient classical long-time evolution and reconstruction of $G_A(t)$.
}
    \label{fig:scheme}
\end{figure*}

\subsection{Hybrid estimation of dynamical correlator}\label{sec:subspacealgo}

We assume that a weakly simulatable classical approximation of $\ket{\psi_A}$ is available, whose Born amplitude can be sampled efficiently~\cite{Nest2008ClassicalSO}. 
This approximation can be obtained in a variety of different ways: as an example a neural network quantum state (NQS~\cite{Carleo2017,misery2025}, foundation NQS~\cite{Rende2025Foundation}) or tensor network representations optimized with protocols such as VMC or DMRG~\cite{Schollwock_2011,Collura_2024}.
While we restrict ourselves to classical preparation of the ground state, this could even be a state prepared on the quantum device, as it would allow sampling from the Born distribution, or a hybrid classical ground state obtained with sample-based quantum algorithms.


The Green’s function will be reconstructed by classically sampling configurations from $|\psi_A|^2$ and evolving each corresponding product state on a quantum processor. 
This circumvents the costly ground-state preparation on hardware and the fully classical simulation of time dynamics. 
The central quantity to estimate is the Loschmidt amplitude of the perturbed state,
\begin{equation} \label{eq:loschmidtecho}
\mathcal{L}_A = \langle \psi_A | \hat{U}(t) | \psi_A \rangle\,.
\end{equation}
Below, we discuss two different strategies to compute the Lochsmidt amplitude. 
\textit{We ultimately rely on the second one for our algorithm, but find that understanding why the first strategy cannot work helps understand the motivation behind our approach.}

\paragraph{Sampling Strategy} --- The Lochsmidt amplitude could be recast into a classical expectation value over the distribution $|\psi_A(\mathbf{x})|^2$,
\begin{multline}
\mathcal{L}_A = \sum_x |\psi_A(\mathbf{x})|^2\frac{\bra{ \psi_A} \hat{U}(t)\ket{\mathbf{x}}}{\braket{\psi_A}{\mathbf{x}}} = \\ = \mathbb{E}_{\mathbf{x}\sim|\psi_A(\mathbf{x})|^2}\left[\frac{\bra{ \psi_A}  \hat{U}(t) \ket{\mathbf{x}}}{\braket{\psi_A}{\mathbf{x}}}\right]\,,
\end{multline}
where $\textbf{x} \in \{ 0,1 \}^{2M}$ indicates a state in the basis of Slater determinants
\begin{equation}
\ket{\mathbf{x}} = \prod_{p=0}^{2M} \left(\hat{c}^\dagger_{p}\right)^{x_{p}} \ket{\varnothing}\,  \, \, \, \, x_p \in \{ 0,1 \} 
\label{eq:slater_bts}
\end{equation}
with $\ket{\varnothing}$ denoting the vacuum state, and $\psi_A(\mathbf{x}) = \langle \mathbf{x} | \psi_A \rangle$ its associated wavefunction amplitude. 
Notably, the variance of this estimator is bounded as
\begin{equation}
\text{Var}[\mathcal{L}_A] = 1 - |\mathcal{L}_A|^2 \leq 1\,,
\end{equation}
implying that it can be estimated with a finite number of classical samples.

While $\braket{x}{\psi_A}$ can be efficiently computed from the classical ansatz, $\bra{\psi_A}U(t)\ket{\mathbf{x}}$ cannot, and would need to be estimated by sampling from the quantum hardware.
For example, it could be obtained from the conditional distribution $p(\mathbf{y}|\mathbf{x})=\abs{\bra{\mathbf{y}}U(t)\ket{\mathbf{x}}}^2$ measured on the quantum hardware (see \Cref{app:alternative}). However, the variance of this estimator scales as the size of the support of $p(\mathbf{y}|\mathbf{x})$, namely
\begin{equation}
\sum_y \chi\left( |\langle \mathbf{y} |  \hat{U}(t) | \mathbf{x} \rangle|^2 > 0 \right).
\end{equation}
where $\chi$ denotes the indicator function. In generic ergodic systems this support grows exponentially with system size, $\sim 2^{2M}$, where again $M$ is the size of spatial orbitals basis set. However, in systems exhibiting dynamical constraints, integrability, or Hilbert space fragmentation, the evolution of $\ket{\mathbf{x}}$ remains confined to a much smaller sector. 
In such sparse regimes, the variance remains bounded and the estimation of $\langle\psi_A|\hat U(t)|\mathbf{x}\rangle$ becomes feasible at polynomial cost.


\paragraph{Subspace Evolution} --- This is the procedure we will employ for the rest of the manuscript.
Instead of estimating  $\bra{\psi_A} \hat{U}(t)\ket{\mathbf{x}}$ with the quantum hardware, we use the device only to \emph{discover} the small portion of Hilbert space that each product state explores at short times and then  classically compute the evolution in the small subspace.

First, for every configuration $x$ drawn from the classical distribution $|\psi_A(\mathbf{x})|^2$, we prepare the product state $\ket{\mathbf{x}}$ on the quantum processor, apply the propagator $ \hat{U}(t)$ for a short, fixed window $t\in[0,t_{\max}]$ with a shallow circuit, and repeatedly measure in the computational basis.  The collection of outcomes $\{\mathbf{y}_t\}$ produced by this \emph{exploratory} evolution reveals the small set of computational basis states that actually acquire weight, and hence defines an empirical support (or \emph{local subspace})

\begin{multline}
\notag \mathcal{S}_x = \big\{ \mathbf{y} \in \{0,1\}^{2M} \, \big| \, \mathbf{y} \sim |\bra{\mathbf{y}}\hat{U}(t)\ket{\mathbf{x}}|^2 ,\, t\in[0,t_{\rm max}]\big\}.
\end{multline}

In the second stage we project the Hamiltonian onto this subspace, \begin{equation}
\hat{H}_{\rm proj}^{(x)}=\hat{P}_{\mathcal{S}_x}\,\hat{H}\,\hat{P}_{\mathcal{S}_x}
\end{equation}
where $\hat{P}_{\mathcal{S}_x} = \sum_{\mathbf{y} \in \mathcal{S}_x} \dyad{\mathbf{y}}$, and classically compute the evolution $e^{-iH_{\rm proj}^{(x)}t}$ to arbitrarily long times with exact diagonalization of the small projected Hamiltonian.
Because $\dim(\mathcal{S}_x) \ll 2^{2M}$ for typical physical (and in this case molecular) Hamiltonians, this hybrid workflow exchanges deep, noise-sensitive quantum circuits for shallow sampling and small-matrix exponentiation, yielding an efficient and variance-controlled estimate of the Loschmidt amplitude.

In the spirit of subspace methods, we approximate the quantum evolution by projecting it onto the subspace $\mathcal{S}_x$. By inserting a resolution of the identity within $\mathcal{S}_x$, the Loschmidt amplitude can be rewritten as
\begin{equation}
\mathcal{L}_A(t) \approx \sum_\mathbf{x} \sum_{\mathbf{y} \in \mathcal{S}_x}  \psi_A^*(\mathbf{y}) \langle \mathbf{y}| e^{-i\hat{H}_{\rm proj}t}| \mathbf{x}\rangle \psi_A(\mathbf{x}).
\end{equation}
Therefore, for each $\ket{\mathbf{x}}$ we reconstruct the dynamics and Loschmidt amplitude with the approximate quantum states $\ket{\phi_x^{\rm proj}(t)} = e^{-i\hat{H}_{\rm proj}t}| \mathbf{x}\rangle $.
We recap the full method in Algorithm \ref{alg:hybrid} and schematically in Fig. \ref{fig:scheme}.

\begin{figure}[t]
  \begin{algorithm}[H]
    \caption{Hybrid Subspace Algorithm}
    \label{alg:hybrid}
    \begin{algorithmic}[1]
      \State Prepare ground state $\ket{\psi_0}$ classically 
      \State Apply $A$ to obtain $\ket{\psi_A} = A \ket{\psi_0}$
      \State Sample configurations $\mathbf{x} \sim |\psi_A(\mathbf{x})|^2$
      \For{each configuration $\mathbf{x}$}
          \State Prepare $\ket{\mathbf{x}}$ on quantum device
          \For{each $t = j\Delta t,\ j = 0, \dots, n_j$}
              \State Evolve $\ket{\phi_x(t)} = \hat{U}(t)\ket{\mathbf{x}}$
              \State Measure $\ket{\phi_x(t)}$ to obtain $\mathbf{y}_t$
          \EndFor
          \State Define subspace $\mathcal{S}_x$ from all collected $\{\mathbf{y}_t\}$
          \State Simulate projected dynamics in $\mathcal{S}_x$ classically
      \EndFor
      \State Reconstruct $\mathcal{L}_A$ from projected dynamics
    \end{algorithmic}
  \end{algorithm}
\end{figure}

The validity of the above approximations rests on a few key assumptions.
First, the projected evolution faithfully reproduces the true dynamics only if the subspace $\mathcal{S}_x$ captures the dominant contributions of the time-evolved state $\ket{\phi_x(t)}$.
This requires that the quantum evolution remains approximately confined within $\mathcal{S}_x$: significant leakage outside it degrades accuracy.
Second, the dimension of $\mathcal{S}_x$ must remain small enough to allow for efficient classical simulation of the projected Hamiltonian $H_{\rm proj}$.
If $\mathcal{S}_x$ grows too large, the subspace restriction offers no computational advantage.
We remark that these are the same assumptions as in other sample-based methods that prepare time-evolved quantum states on quantum devices to sample from them \cite{yu2025sqkd,piccinelli2025}.

Fortunately, in many relevant physical systems, the action of $\hat{U}(t)$ on a basis state $\ket{\mathbf{x}}$ produces a state $\ket{\phi_x(t)}$ that remains sparse in the computational basis: only a small number of configurations acquire appreciable amplitude. 
This effective sparsity often arises due to Hilbert space fragmentation, where the structure of the Hamiltonian dynamically restricts the evolution to disconnected or weakly connected sectors of the Hilbert space.

This sparsity is what ultimately makes both quantum sampling and classical projection within $\mathcal{S}_x$ computationally feasible.
Consequently, short-time quantum evolutions suffice to identify the dominant configurations, which
define an effective subspace in which the dynamics can be extrapolated to longer times using purely classical computations. 

This is particularly advantageous in applications such as molecular spectroscopy, where we are interested in long-time observables like excitation energies.
In these cases, the energy gap between the ground and excited states is typically small, meaning that relevant spectral features are encoded in slow oscillations of the correlation function $G_A(t)$.
As a result, accurate spectral estimates can be obtained even when the quantum evolution is available to modest time intervals with shallow circuits.

A final important aspect concerns implementing short-time evolution on hardware, i.e., preparing $\ket{\phi_x(t)}$. 
Although time evolution can, in principle, be implemented efficiently on a quantum device \cite{lloyd1996, Berry_2015, Tacchino_2019}, several techniques have been developed in recent years to make it feasible on contemporary, error-prone hardware. 
This is particularly relevant for molecular systems, where the number of Hamiltonian terms scales as $O(M^4)$, and their mapping onto quantum devices can generate deep, non-local quantum gates, making it challenging to implement even a single Trotter step \cite{mikkelsen2025qscite}.
However, very recently a new technique has been proposed that leverages the quantum drift (qDRIFT \cite{Campbell_2019}) randomized compilation strategy to execute large scale and provably convergent quantum chemistry experiments on quantum devices \cite{piccinelli2025}.
We envision that combining our proposed technique for computing dynamical correlation functions with the sample-based qDRIFT protocol (SqDRIFT) could enable the large-scale determination of dynamical correlators.

\begin{figure*}
    \centering
    \includegraphics[width=\linewidth]{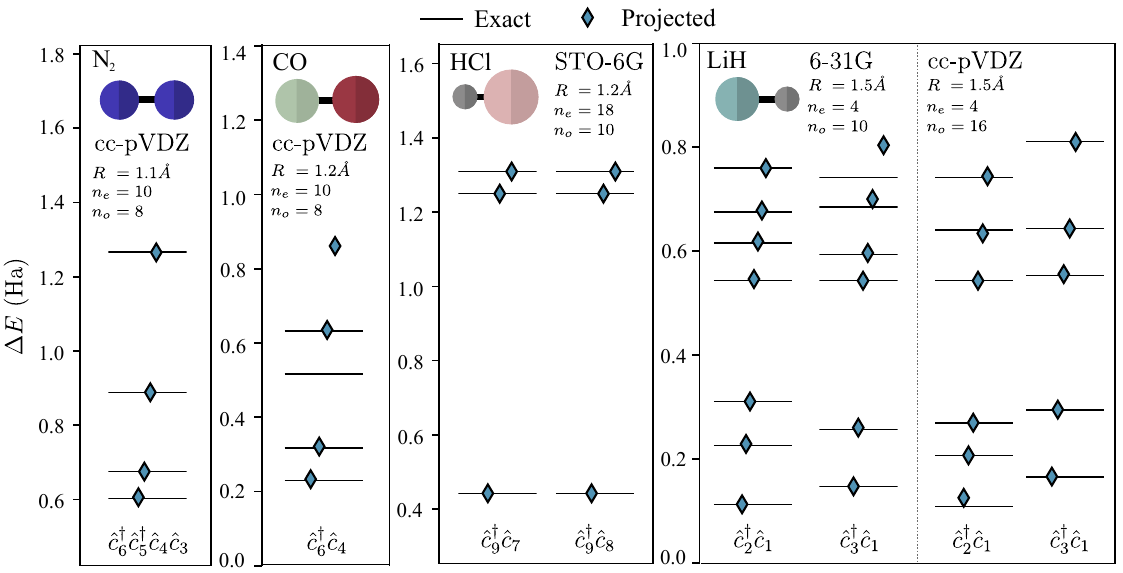}
    \caption{
    \textbf{Energy spectra of small molecular systems.}
We compare excitation energies of various molecular systems computed using the hybrid classical–quantum method (blue diamonds), against exact diagonalization results (black lines). 
All systems are small enough to allow full diagonalization of the Hamiltonian, which serves as a reference for validating the projected dynamics.  
For each system, we report the number of spin-orbitals ($n_o$), number of electrons ($n_e$), interatomic distance $R$ (in \AA) and basis set. 
For each molecule, the excitation operator $\hat{A}$ used to define the perturbed state $\ket{\psi_A} = \hat{A} \ket{\psi_0}$ is designed to selectively target low-energy excitations. 
These include single and double excitations, such as $\hat{A} = c^\dagger_6 c^\dagger_5 c_4 c_3$ for N$_2$ , $\hat{A} = c^\dagger_9 c_7$ and $\hat{A} = c^\dagger_9 c_8$ for HCl and, and $\hat{A} = c^\dagger_6 c_4$ for CO. 
In the LiH systems, different perturbations are tested across basis sets: with 6-31G uses $\hat{A} = c^\dagger_2 c_1$, with cc-pvdz uses the same operator, while 6-31G (g) and cc-pvdz use $\hat{A} = c^\dagger_3 c_1$.
The agreement between projected dynamics and exact results across all panels confirms the accuracy and flexibility of the method.
}
    \label{fig:all_molecules_peaks}
\end{figure*}

\section{Results}\label{sec:result}
We validate our hybrid classical–quantum approach by applying it to a set of molecular systems of increasing complexity. 
The first benchmarks, presented in \cref{sec:result_ed}, focuses on small molecules, for which exact solutions are available.
The second set of benchmarkes, presented in \cref{sec:results_tn}, focuses on larger molecules that we tackle with Tensor Network algorithms.
This allows us to directly assess the accuracy of our method by comparing the projected subspace dynamics against results obtained from full diagonalization. 

Ultimately, we showcase the validity of our methods and that a realistic implementation on quantum hardware is feasible.

\subsection{Exact Diagonalization}
\label{sec:result_ed}
Results in this section are obtained by performing exact diagonalization of the Hamiltonian to find the ground-state and time-evolution.

Fig.~\ref{fig:all_molecules_peaks} presents the excitation spectra for several representative molecules (including N\textsubscript{2}, HCl, CO, and LiH) using a variety of basis sets and orbital configurations. Each panel compares the exact excitation energies (black lines) with the spectrum reconstructed from the projected evolution (blue diamonds), where energy levels are extracted from the long-time behavior of the correlation function.
The excellent agreement observed across all systems, particularly in the low-energy sector, demonstrates that our method can faithfully reproduce the relevant spectral features, even when the subspaces used in the projection remain of moderate size. 
This confirms both the accuracy and the computational efficiency of the hybrid approach. We highlight that the agreement is robust for the lowest excited states, which are often then most technologically-relevant ones.

We kept only those excitations for which the Fourier amplitude satisfies $|\tilde{G}(\omega)| > 5 \times 10^{-3}$, ensuring that the extracted peaks are statistically significant and not artifacts of noise. 
Notice, however, that since the absolute amplitude of the peaks in the projected spectra is not directly resolved (see \Cref{app:fourierspectra}), some excitations may fall below the threshold even though they are physically present in the projected dynamics.
In \Cref{app:fourierspectra}, we show the full Fourier spectra of the molecules displayed in Fig.~\ref{fig:all_molecules_peaks}. 
As a result, certain spectral features that are present in the projected dynamics may be inadvertently excluded or included in the final excitation spectrum, depending on whether their amplitude lies just below or above the fixed threshold. 
This effect is visible, for example, in the case of the CO molecule shown in Fig.~\ref{fig:all_molecules_peaks}(d).

To construct the subspaces $\mathcal{S}_x$ needed for the projected evolution, we first perform short-time quantum simulations for each sampled configuration $\mathbf{x}$.
Specifically, we evolve the state $\ket{\mathbf{x}}$ under the full unitary propagator $\hat{U}(t)$ and perform 4096 projective measurements in the computational basis at each discrete time step $t = 0, 1, 2, \ldots, 10$. 
The union of all configurations observed during this initial window defines the effective subspace $\mathcal{S}_x$ in which the dynamics will be later confined. 
This subspace captures the region of the Hilbert space actually explored by the evolution, and its size reflects the sparsity structure induced by the system Hamiltonian and the choice of initial state. 
Once $\mathcal{S}_x$ is built, we proceed to simulate the time evolution within the subspace entirely classically, extending it to much longer times over the interval $t \in [-1000, 1000]$. 

While we don't report it in the main text, in \cref{app:sample_scaling} we present a detailed analysis of convergence with respect to the number of quantum measurements, where we also discuss how the accuracy threshold is ultimately set by the finite time window of the classical propagation.

We remark that this separation of timescales (using shallow quantum circuits for subspace discovery and classical computation for extended dynamics) allows us to extract high-resolution spectra without requiring deep quantum resources. 
Notably, reaching such long time scales can be computationally demanding even for classical techniques, due to the accumulation of numerical errors in the time discretizations or the growth of entanglement in exact or approximate methods.
By working in a reduced and dynamically relevant subspace, our method mitigates this challenge and enables accurate long-time evolution at lower cost.

In Table~\ref{tab:molecules}, we summarize the systems analyzed in our benchmarks. 
For each molecule, we report the number of orbitals, number of electrons, bond length, basis set, the excitation operator $\hat{A}$ used to perturb the ground state, and the maximum size of the corresponding projected subspace. 
Note that different perturbations can lead to variations in the subspace size (even for the same molecule).
Nonetheless, as shown by comparison with the excited-state energies, accurate results are still achieved.

\begin{table}[t]
\centering
\begin{tabular}{lccccc c}
\toprule
Molecule & $n_o$ & $n_e$ & $R$ & Basis & Perturbation & Max ${\rm dim}\left[ \mathcal{S}_x \right]$ \\
\midrule
N$_2$    & 8     & 10    & 1.1       & cc-pvdz   & $\hat{c}^\dagger_6 \hat{c}^\dagger_5 \hat{c}_4 \hat{c}_3$ & 341  \\ 
HCl      & 10    & 18    & 1.2       & STO-6G    & $\hat{c}^\dagger_9 \hat{c}_7$                 & 8    \\ 
HCl      & 10    & 18    & 1.2       & STO-6G    & $\hat{c}^\dagger_9 \hat{c}_8$                 & 8    \\ 
CO       & 8     & 10    & 1.2       & cc-pvdz   & $\hat{c}^\dagger_6 \hat{c}_4$                 & 821  \\ 
LiH      & 10    & 4     & 1.5       & 6-31G     & $\hat{c}^\dagger_2 \hat{c}_1$                 & 331  \\ 
LiH      & 16    & 4     & 1.5       & cc-pvdz   & $\hat{c}^\dagger_2 \hat{c}_1$                 & 1421 \\ 
LiH      & 10    & 4     & 1.5       & 6-31G     & $\hat{c}^\dagger_3 \hat{c}_1$                 & 298  \\ 
LiH      & 16    & 4     & 1.5       & cc-pvdz   & $\hat{c}^\dagger_3 \hat{c}_1$                 & 1007 \\ 
\bottomrule
\end{tabular}
\caption{Molecular configurations and simulation parameters used in our benchmark tests. For each system, we report the number of spin-orbitals ($n_o$), number of electrons ($n_e$), interatomic distance $R$ (in \AA), basis set, the perturbation operator used to excite the system, and the maximum dimension of the effective basis subspaces.}

\label{tab:molecules}
\end{table}

\subsection{Tensor Networks}
\label{sec:results_tn}

\begin{figure}[t]
  \centering
\includegraphics[width=\linewidth]{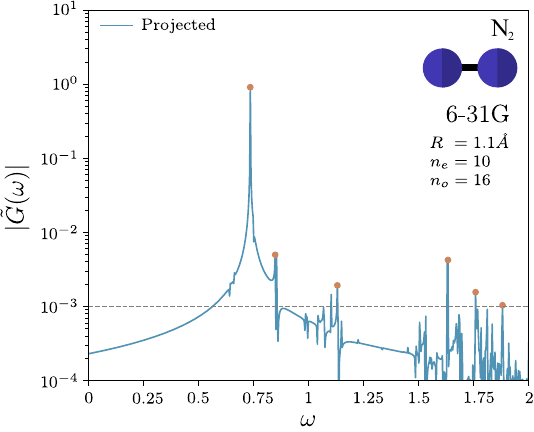}
  \caption{\textbf{$\text{N}_2$ dynamical correlation using MPS.}
  Excitation spectrum for the $\text{N}_2$ molecule in the 6-31G basis set with frozen core approximation.
  The ground state is represented with a MPS state optimized using the DMRG algorithm, as described in \cref{sec:results_tn}.
  The peaks are at the following energies $\Delta E =0.735, 0.848, 1.131, 1.633.$
  }
  \label{fig:N2_largemolecule}
\end{figure}

To explore the scalability of our method beyond sizes accessible to exact diagonalization, we complement our benchmarks with some tensor network calculations. 
In particular, we employ matrix product states (MPS) to simulate both the ground state and the quantum dynamics when exact methods are no longer feasible. 

We recall that an MPS is a structured variational ansatz for many-body quantum states. 
In the case of molecules, the wavefunction $\ket{\Psi}$ of a system with $M$ orbitals can be expressed as
\begin{equation}
\ket{\Psi} = \sum_{\{x_i\}} \Tr{B_1^{(x_1)} B_2^{(x_2)} \cdots B_{M}^{(x_{M})} }\ket{x_1 x_2 \ldots x_{\rm 2M}},
\end{equation}
where $x_i = 0,1,2,3$ labels the local basis state at orbital $i$, corresponding respectively to an empty site, spin-up occupation, spin-down occupation, and double occupation. For a given $x_i$ each tensor $B_i^{(x_i)}$ is a matrix of size at most $m \times m$ (with boundary dimensions adjusted at the ends). The bond dimension $m$  controls the number of variational parameters and sets the maximum amount of entanglement that can be captured across any bipartition of the system.

In our implementation, we employ the DMRG to variationally optimize the ground state $\ket{\psi_0}$ within the MPS ansatz. The perturbed state $\hat{A}\ket{\psi_0}$ (with the same amount of entanglement as $\ket{\psi_0}$) is then used to generate samples in the computational basis. For each sampled configuration $\ket{x_1 \cdots x_{M}}$, corresponding to a product state with bond dimension $m = 1$, we perform time evolution using the time-dependent variational principle (TDVP). From the time-evolved states, we extract the configurations that contribute significantly to the dynamics and use them to construct the effective subspaces $\mathcal{S}_x$. If $\log_2[{\rm dim}(\mathcal{S}_x)] \approx 20\sim 25$ we can perform efficiently with a classical sparse matrix algorithm the quantum evolution up to large times.

We consider an $\text{N}_2$ molecule with 16 spin-orbitals and 10 electrons, using the 6-31G basis set. The ground state was computed via DMRG with bond dimension $m = 512$, yielding an energy error of approximately $10^{-5}$. We perturbed the ground state with the operator $\hat{A} = c^\dagger_6 c^\dagger_5 c_4 c_3$ and sampled 562 basis states using $10^6$ measurements. These samples were then evolved using the TDVP algorithm and projected at times $t = 0, 1, 2, 3$, each with $10^6$ additional samples, the maximum bond dimension of the evolution is $m=64$ and a cutoff of $10^{-8}$.  In order to keep the most important states we considered a projection in $\bigcup_x \mathcal{S}_x$.  This procedure effectively reduced the dimensionality of the Hilbert space relevant for the dynamics from approximately $2^{24}$ to $2^{16}$. In Fig.~\ref{fig:N2_largemolecule} we show the excitation spectrum, notice that we found energies compatible to the ones in Fig.~\ref{fig:all_molecules_peaks}(a). Further investigations, including simulations of larger molecules and the implementation of the dynamics on quantum hardware, appear especially promising for exploring the scalability and accuracy of this approach.

\section{Conclusions}\label{sec:conclusions}

We have introduced a hybrid classical–quantum algorithm to compute dynamical correlation functions and extract excitation spectra of interacting fermionic systems, with a focus on quantum molecular spectra. 
The novel idea is to combine classical variational ground-state optimization with time-evolution on the quantum hardware, without enforcing a specific classical ansatz structure as has been investigated previously~\cite{Rudolph2023TNPretraining}.
We build a classical approximation of the perturbed state, sample product states from it, and then evolve them using shallow quantum circuits. 
From these short-time evolutions, we construct dynamically relevant subspaces $\mathcal{S}_x$ in which we simulate the full time evolution classically. 
This strategy significantly reduces quantum resource requirements while retaining access to long-time dynamics and high-resolution spectra.

We validated the method on a set of benchmark molecular systems using exact diagonalization, demonstrating excellent agreement in the low-energy excitation spectrum even when the projected subspace is of modest dimension. 
Our results show that the dominant dynamical features can be captured efficiently within small subspaces, thanks to the effective sparsity induced by the structure of the Hamiltonian and the perturbation. 
We further illustrated how this method can be combined with tensor network techniques, such as DMRG and TDVP, to extend its applicability to larger systems where exact diagonalization is no longer feasible.

This framework opens up several promising directions. 
On near-term quantum devices, our protocol provides a practical route to extract spectroscopic observables with reduced circuit depth and minimal overhead.
Once again, we remark the possibility to integrate this protocol with recent advances in sample-based quantum algorithm to scale accurate computation of dynamical properties of strongly correlated quantum matter beyond the reach of exact methods.
On the classical side, the identification of small, dynamically relevant subspaces offers a pathway to overcome time-scale limitations in existing simulation methods. 
Future extensions could include adaptive subspace refinement and applications to other strongly correlated systems such as impurity problems~\cite{Bravyi2017Impurity}.

\section*{Acknowledgments}
We acknowledge the use of ITensors for the tensor network simulations~\cite{itensor}, as well as PySCF~\cite{pyscf}, QuTiP~\cite{qutip}, and NetKet~\cite{netket3:2022,netket2:2019} for various parts of the implementation. 
We acknowledge T. Ayral, M. Rudolph, D. Castaldo, J. Landman E. Kashefi for insightful discussions and Z. Holmes for introducing them. 
F.V. also acknowledges numerous and unforgettable discussions with G. Carleo.
This work was granted access to the HPC resources of IDCS support unit from Ecole polytechnique. 
This work was provided with computing HPC and storage resources by GENCI at IDRIS thanks to the grants 2024-A0170515698 and 2025-AD010916479 on the supercomputer Jean Zay's CPU and V100 partitions.
F.V. and A.S. acknowledge support by the French Agence Nationale de la Recherche through the NDQM project, grant ANR-23-CE30-0018. 
A.S. acknowledges support by Hi!Paris through the Postdoctoral fellowship, grant HiP-2025-07.

\bibliography{bibliography}

\clearpage
\newpage

\appendix
\onecolumngrid

\section{Complete Fourier Spectra Comparison}
\label{app:fourierspectra}

\begin{figure*}[h]
    \centering
    \includegraphics[width=\linewidth]{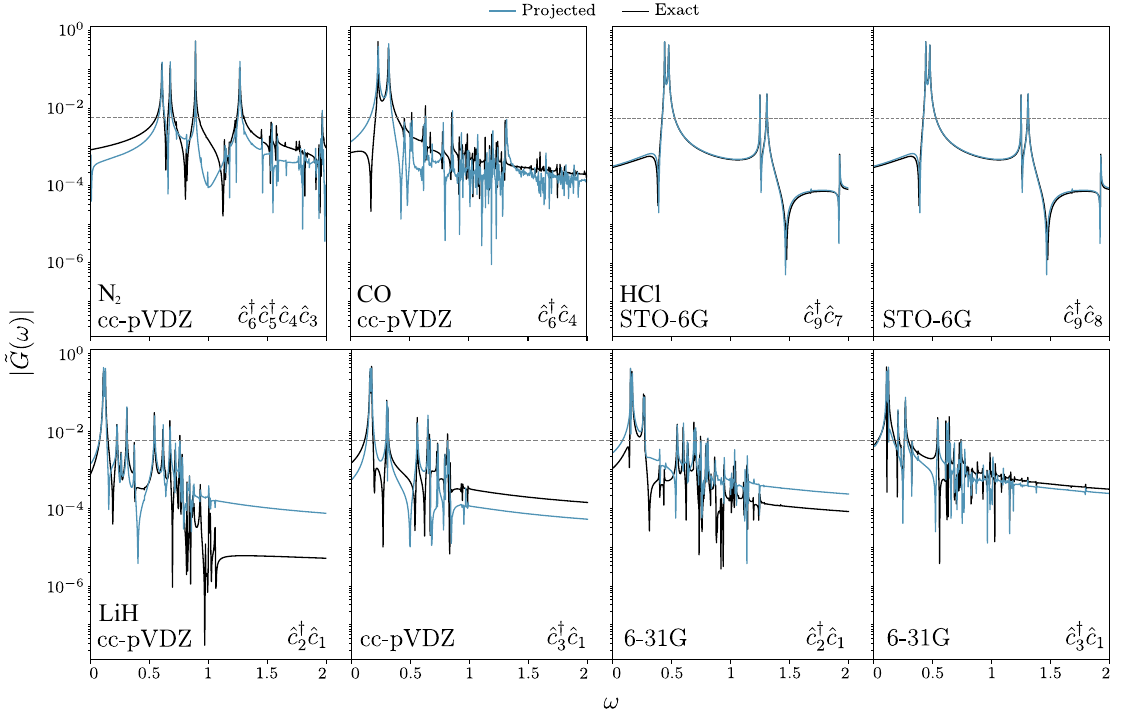}
    \caption{
    \textbf{Dynamical correlation functions of small molecules.}
    Absolute value of the dynamical correlation function, comparison of the exact (black line) with the projected one (blue lines).
    The grey dashed line at $|\tilde{G}(\omega)| = 5 \times 10^{-3}$ indicates the threshold used to filter out peaks.
    Parameters as in Fig. \ref{fig:all_molecules_peaks}. 
    }
    \label{fig:all_molecules_fourier}
\end{figure*}

To support the spectral analysis presented in the main text, we provide here the complete Fourier-transformed correlation functions for the set of molecules analyzed in Fig.~\ref{fig:all_molecules_peaks}. 
We display in Fig.~\ref{fig:all_molecules_fourier} the absolute value of the Fourier-transformed correlation function $|\tilde{G}(\omega)|$ for a representative set of molecules (see Fig.~\ref{fig:all_molecules_peaks} for details), comparing the exact (black) and projected (blue) spectra. 
In all cases, we observe that the projected spectra accurately reproduce the dominant features of the exact signal, particularly in the low-frequency region ($\omega < 1 {\rm Ha}$), where the most physically relevant excitations reside.

For systems such as N\textsubscript{2}  and CO, the agreement is striking across a wide frequency range, with nearly all major peaks correctly recovered. 
HCl also shows excellent agreement, with the projected dynamics capturing the multiplicity and spacing of low-lying transitions. 
In LiH configurations, although the high-frequency part of the spectrum becomes noisier (especially for large orbital sets) the location and relative strength of the main peaks are still well reproduced. 
The horizontal dashed line at $|\tilde{G}(\omega)| = 5 \times 10^{-3}$ marks the empirical threshold used to filter out spurious peaks.

Overall, the results confirm that the projected evolution faithfully reconstructs the spectral structure, especially for the low-energy sector.

\section{Scaling with Number of Quantum Samples}
\label{app:sample_scaling}

\begin{figure}
    \centering
    \includegraphics[width=\linewidth]{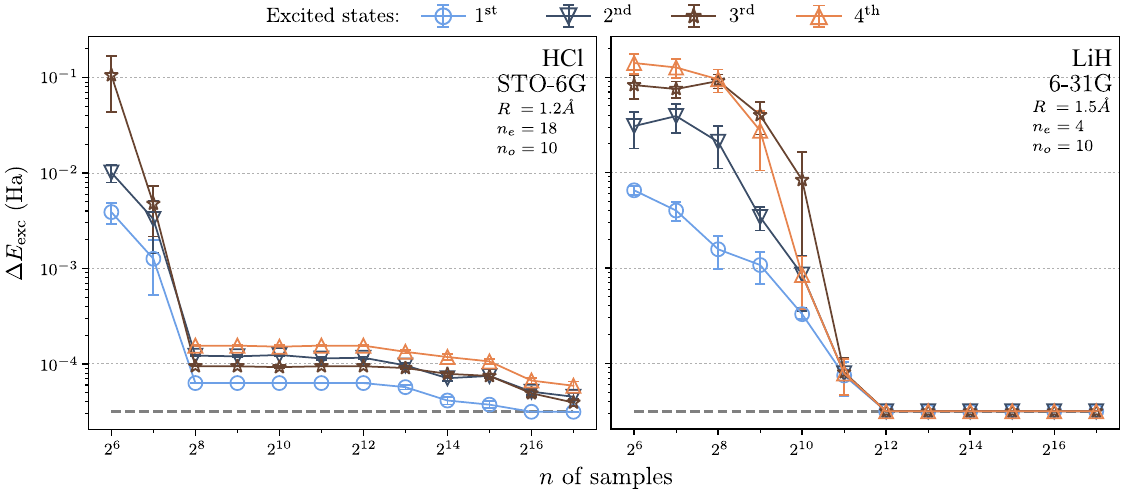}
    \caption{
    \textbf{Excited states accuracy scaling with number of samples.}
    Accuracy of the extracted excitation energies as a function of the number of quantum samples used to construct the subspace $\mathcal{S}_x$.
    While the behavior depends on the type of system and its size, we observe a transition to very high accuracy with a relatively small number of samples ($\sim 2^{10}$).
    The accuracy is limited by the spectral resolution of the Fourier transform, here indicated as a grey dashed line.
    }
    \label{fig:scalingsamples}
\end{figure}

To quantify the sampling requirements of our hybrid protocol, we examine how the error in the estimated excitation energies decays with the number of quantum samples used to define the subspace $\mathcal{S}_x$. Fig.~\ref{fig:scalingsamples} shows this behavior for two representative molecules. 
We observe rapid convergence: for all excited states shown, the error decays quickly with increasing sample size and stabilizes beyond $2^{10}$ samples. 
Importantly, the residual error plateau reflects the finite time window $\Delta t \in [-T, T]$ used in the classical projected evolution. 
Since spectral resolution is inversely proportional to $T$, longer classical evolutions provide finer energy discrimination.

\section{Alternative Estimator for the Loschmidt Amplitude}
\label{app:alternative}

In this section we sketch an alternative approach to estimating the Loschmidt amplitude $\mathcal{L}_A$. Here, we express explicitly the normalization of the wavefunctions. Our starting point is the  expression
\begin{equation}
  \mathcal{L}_A = \sum_{x} \frac{|\psi_A(x)|^2}{\langle \psi_A|\psi_A\rangle}\,\ell(x),
\end{equation}
where we defined the local estimator
\begin{equation}
  \ell(x) = \frac{\langle \psi_A|U(t)|x\rangle}{\psi^*_A(x)}.
\end{equation}

We then introduce a complete basis $\{|y\rangle\}$ to expand $\ell(x)$
\begin{align}
  \ell(x) &= \frac{\langle \psi_A|U(t)|x\rangle}{\psi^*_A(x)}
  = \sum_y \frac{\langle \psi_A|y\rangle\,\langle y|U(t)|x\rangle}{\psi^*_A(x)} \nonumber\\[1mm]
  &= \sum_y \frac{\psi^*_A(y)}{\psi^*_A(x)}\,\langle y|U(t)|x\rangle\,.
\end{align}

Noticing that $|\langle y|U(t)|x\rangle|^2$ is the transition probability to the bit string $\ket{y}$ from the evolved state $\ket{\phi_x(t)}=U\ket{x}$ we have
\begin{equation}
  \ell(x) = \sum_y \frac{\psi^*_A(y)}{\psi^*_A(x)}\,\frac{|\langle y|U|x\rangle|^2}{\langle x|U(t)|y\rangle}\,.
\end{equation}

This formulation suggests a sampling strategy: for each fixed $x$, sampled from the probability distribution
\begin{equation}
  p(x)=\frac{|\psi_A(x)|^2}{\langle \psi_A|\psi_A\rangle}\,,
\end{equation}
we sample $y$ from the conditional distribution
\begin{equation}
  p(y\mid x)=|\langle y|U|x\rangle|^2\,.
\end{equation}
We then define the estimator
\begin{equation}
  f(y,x) = \frac{\psi^*_A(y)}{\psi^*_A(x)}\frac{1}{\langle x|U(t)|y\rangle}\,,
\end{equation}
so that our target becomes
\begin{equation}
  \mathcal{L}_A = \mathbb{E}_{(x,y)\sim p(y\mid x)p(x)}\left[f(y,x)\right]\,.
\end{equation}
A brief calculation reveals, however, that the variance of this estimator behaves as
\begin{align}
  \mathrm{Var}[\mathcal{L}_A] &= \sum_{x,y} p(x)p(y\mid x)\frac{p(y)}{p(x)p(y\mid x)}- |\mathcal{L}_A|^2\,,\notag \\
  &=2^{2M}-|\mathcal{L}_A|^2
\end{align}
which grows exponentially with the number of qubits $2M$. This is particularly concerning when $ |\langle x|U(t)|y\rangle|^2 $ is nonzero in all of the Hilbert space. The variance of the estimator will scale with the full dimension $2^{2M}$. On the other hand if the dynamics explores only a small subset of the Hilbert space we have that\begin{align}
    \notag    {\rm Var}[\mathcal{L}_A] &= \sum_{y} p(y) \sum_x\chi\biggl(p(y\mid x)>0\biggr) - |\mathcal{L}_A|^2\\ &= \sum_{y} p(y){\dim \mathcal{S}_y}-|\mathcal{L}_A|^2
\end{align}
where $\chi$ is the indicator function and $\mathcal{S}_y$ is the subspace explored by the dynamics starting from $y$.  
Notice that if we consider $t=0$ we obtain $p(y|x) = \delta(x-y)$, thus ${\dim }\,\mathcal{S}_y=1$ and ${\rm Var}[\mathcal{L}_A]=1-|\mathcal{L}_A|^2$. The variance is therefore determined by the size of this region that supports $\ket{\phi_x(t)}=\hat{U}(t)|x\rangle$.

\end{document}